\documentclass[11pt]{article}
 \usepackage{graphicx,amsfonts,amsmath}

 \renewcommand{\title}[1] {%
 \begingroup\begin{center}\vspace{0.0cm}\bf\Large
 \addtolength{\baselineskip}{1mm} #1 \end{center}\endgroup}

 \renewcommand{\author}[1] {%
 \begingroup\begin{center}\vspace{0.2cm}\bf #1 \vspace{0.2cm}
 \end{center}\endgroup}

  \newcommand{\address}[1] {%
 \begingroup\begin{center} #1 \end{center}\endgroup}

\begin{document}
 \title{Spin matrix elements in 2D Ising model \\ on the finite lattice}
 \author{A. I. Bugrij$^{\; *}$, O. Lisovyy$^{\;*,\;\dag}$}
 \address{
  $^{*\;}$Bogolyubov Institute for Theoretical Physics \\
  Metrolohichna str., 14-b, Kyiv-143, 03143, Ukraine \vspace{0.2cm} \\
  $^{\dag\;}$ Laboratoire de Math\'ematiques et Physique Th\'eorique CNRS/UMR 6083,\\
  Universit\'e de Tours, Parc de Grandmont, 37200 Tours, France}
  \date{}

 \begin{abstract}
 We present explicit formulas for all spin matrix elements in the 2D
 Ising model with the nearest neighbor interaction  on the finite periodic
 square lattice.
 These expressions generalize the known  results
 \cite{karowski, bugrij, fonseca} (coincide with them in the
 appropriate limits) and fulfill the test of straightforward transfer matrix
 calculations for finite $N$.
 \end{abstract}
 \section{Eigenvalues and eigenvectors of the transfer
 matrix}
 It is well-known (see \cite{huang, mccoy}) that the spectrum of $2^N\times 2^N$
 transfer matrix, corresponding to Ising model on the periodic square
 lattice, consists of two sets:
 $$
 \lambda=(2s)^{N/2}\exp\left\{\frac12\left(\pm\gamma\left(0\right)
 \pm\gamma(2\pi/N)\pm\ldots
 \pm\gamma(2\pi-2\pi/N)\right)\right\},\eqno(1)
 $$
 $$
 \lambda=(2s)^{N/2}\exp\left\{\frac12(\pm\gamma(\pi/N)
 \pm\gamma(3\pi/N)\pm\ldots
 \pm\gamma(2\pi-\pi/N))\right\},\eqno(2)
 $$
 where $s=\sinh 2\mathcal{K}$ and $\mathcal{K}$ is the Ising
 coupling constant. The function $\gamma(q)$ is defined as the
 positive root of the equation
 $$ \cosh\gamma(q)=s+s^{-1}-\cos q.$$
 which is the lattice analog of the relativistic energy dispersion law.
 The number of minuses in (1) is even in ferromagnetic ($s>1$) and
 odd in paramagnetic ($0<s<1$) phase, while
 the number of minuses in (2) is even in both phases. The eigenvalues (1)
 (or (2)) correspond to eigenvectors that are odd (resp. even) under
 spin reflection.

 The notation and terminology, introduced in \cite{fonseca} for the
 analysis of continuum limit, are also very convenient on the lattice.
 In what follows, odd and even eigenvectors of the Ising transfer matrix
 will be interpreted as multiparticle states from the Ramond and
 Neveu-Schwartz sector. Quasimomenta of
 R--particles can be equal to $\frac{2\pi}{N}j$ ($j=0,1,\ldots ,N-1$),
 while for NS--particles they take on the values
 $\frac{2\pi}{N}\left(j+\frac12\right)$
 ($j=0,1,\ldots ,N-1$). Each eigenstate consists of particles of
 only one type, and their quasimomenta must be different.

 We will denote by $|p_1,\ldots ,p_K\rangle_{NS(R)}$ the normalized
 eigenstate, containing particles with the momenta $p_1,\ldots ,p_K$.
 Since R--sector in paramagnetic phase contains the state $|0\rangle_R$
 (one particle with zero momentum), it will be convenient to denote
 NS and R vacua by $|\emptyset\rangle_{NS}$ and $|\emptyset\rangle_{R}$.
 The goal of the present paper is to find  matrix elements
 $_{NS}\langle p_1,\ldots,p_K|\sigma|q_1,\ldots,q_L\rangle_{R}$ of the
 Ising spin $\sigma$ in the described basis of normalized eigenstates.
 (R--R and
 NS--NS matrix elements vanish due to $\mathbb{Z}_2$-symmetry of the
 model).
 \section{Lattice form factors and scaling limit}
 All $n$-point correlation functions in the Ising model on the cylinder
 and torus can be easily expressed via spin matrix elements. However,
 known results were obtained in rather inverse way.
 At the first stage, 2-point functions are
 expressed through the determinants of certain Toeplitz
 matrices with a size that depends on the separation of correlating
 spins.
 To extract the analytic dependence on the distance from
 these representations, a lot of further work was needed \cite{bugrij1}.
 The final answer \cite{bugrij,we} allows to calculate squared
 form factors on the cylinder (on the infinite lattice the above program was realized
 earlier in \cite{montroll,palmer,yamada}):
 $$ {\bigl|_{NS}\langle\emptyset|\sigma|q_1,\ldots
      ,q_L\rangle_R\bigr|}^{\, 2}=\xi\, \xi_T
  \prod\limits_{j=1}^{L}\frac{e^{-\nu(q_j)}}{N \sinh\gamma(q_j)}\prod\limits_{1\leq
    i<j\leq L}\left(\frac{\sin
      \frac{q_i-q_j}{2}}{\sinh\frac{\gamma(q_i)+\gamma(q_j)}{2}}
 \right)^2 .\eqno(3)$$
 Here $\xi=\bigl|1-s^{-4}\bigr|^{1/4}$, quasimomenta have discrete
 R--values and
 cylindrical parameters $\xi_T$, $\nu(q)$ are given by
 $$\ln\xi_{T}=\frac{N^{2}}{2\pi^{2}}\int\limits_{0}^{\pi}\int\limits_{0}^{\pi}
 \frac{dp\; dq\, \gamma'(p)\gamma'(q)}{\sinh\left(N\gamma(p)\right)\sinh\left(N\gamma(q)\right)}
 \ln\left|\frac{\sin\left((p+q)/2\right)}{\sin\left((p-q)/2\right)}\right|,\eqno(4)$$
 $$\nu(q)=\frac{1}{2\pi}\int\limits_{-\pi}^{\pi}\frac{dp\, \sinh \gamma(q) }
 {\cosh\gamma(q)-\cos p}
 \ln\coth\left(N\gamma(p)/2\right).\eqno(5)$$
 In the thermodynamic limit $N\rightarrow\infty$ these parameters vanish
 ($\xi_{T}\rightarrow
 1$, $\nu(q)\rightarrow 0$) and (3) transforms into the classical
 formula \cite{palmer,yamada}:
  $$ {\bigl|_{NS}\langle\emptyset|\sigma|q_1,\ldots
      ,q_L\rangle_R\bigr|}^{\, 2}=\xi
  \prod\limits_{j=1}^{L}\frac{1}{\sinh\gamma(q_j)}\prod\limits_{1\leq
    i<j\leq L}\left(\frac{\sin
      \frac{q_i-q_j}{2}}{\sinh\frac{\gamma(q_i)+\gamma(q_j)}{2}}
 \right)^2 ,\eqno(6)$$
 where $\{q_j\}$ can take on
 arbitrary values in the interval $[-\pi,\pi\;]$.

 In the scaling limit, Ising model on the plane was shown to be equivalent to
 a relativistic quantum field theory with two-particle $S$-matrix equal
 to $-1$ (see \cite{smj}). Then it became possible to use the results
 of \cite{karowski} and to calculate all spin matrix elements:
 $$_{NS}\langle p_1,\ldots , p_K|\sigma|q_1,\ldots
      ,q_L\rangle_R=\sqrt{\xi}\; \prod\limits_{i=1}^{K}
\frac{1}{\sqrt{2\pi\omega(p_i)}}
  \prod\limits_{j=1}^{L}\frac{1}{\sqrt{2\pi\omega(q_j)}}
 \;\;F\Bigl(\{p\} \big|\{q\}\Bigr),  \eqno(7)$$
$$F\Bigl(\{p\} \big|\{q\}\Bigr)=\prod\limits_{1\leq i<j\leq K}
\frac{p_i-p_j}{\omega(p_i)+\omega(p_j)}
\prod\limits_{1\leq i<j\leq L}
\frac{q_i-q_j}{\omega(q_i)+\omega(q_j)}
 \prod\limits_{\substack{1\leq i\leq K \\ 1\leq j\leq L}}
\frac{\omega(p_i)+\omega(q_j)}{p_i-q_j}.
\eqno(8)$$
 Here  $\omega(q)=\sqrt{m^2+q^2}$ and the momenta of both type of particles take on arbitrary real
 values. The RHS of (7) is usually written with the factor $i^{[\frac{K+L}{2}]}$, but
 it can be removed by a change of the basis and will be omitted in
 what follows.

 Very recently, Fonseca and Zamolodchikov \cite{fonseca} announced  and promised to
 give a proof of a similar formula for the scaling limit on the cylinder:
$$_{NS}\langle p_1,\ldots , p_K|\sigma|q_1,\ldots
      ,q_L\rangle_R=\sqrt{\xi\,\tilde{\xi}_T}\; \prod\limits_{i=1}^{K}
\frac{e^{\,\tilde{\nu}(p_i)/2}}{\sqrt{\beta\,\omega(p_i)}}
  \prod\limits_{j=1}^{L}\frac{e^{-\tilde{\nu}(q_j)/2}}{\sqrt{\beta\,\omega(q_j)}}
 \;\;F\Bigl(\{p\} \big|\{q\}\Bigr).  \eqno(9)$$
 The overall factor $\tilde{\xi}_T$ and the function $\tilde{\nu}(q)$
 from the leg factors are determined from the scaling limit of (4),
 (5):
 $$\ln\tilde{\xi}_T=\frac{m^2\beta^{\,2}}{2\pi^2}
 \int\limits_{0}^{\infty}\int\limits_{0}^{\infty}
 \frac{dp\; dq\,
   \omega'(p)\omega'(q)}{\sinh\left(\beta\,\omega(p)\right)
 \sinh\left(\beta\,\omega(q)\right)}
 \ln\left|\frac{p+q}{p-q}\right|,\eqno(10)$$
 $$\tilde{\nu}(q)=
 \frac{1}{\pi}\int\limits_{-\infty}^{\infty}\frac{dp \;\,\omega(q)}{p^2+q^2+m^2}
 \;\ln\coth\frac{\beta\omega(p)}{2}.\eqno(11)$$
 Here, $\beta$ denotes the scaled length of the base of the cylinder, NS--momenta are
 quantized as $p_j=\frac{2\pi}{\beta}\, l_j$, $l_j\in
 \mathbb{Z}+\frac12$, while for R--momenta we have $q_j=\frac{2\pi}{\beta}\,
 l_j$ and $l_j\in \mathbb{Z}$.

 \section{General formula}
 On the level of form factors $_{NS}\langle\emptyset|\sigma|\{q\}\rangle_R$
 the expression (9) represents nothing new and can even be proven rigorously --- it is
 simply a particular case of (3). However, this conjecture gives \textit{all}
 matrix elements, though only in the scaling limit. Moreover, the
 structure of this representation is so transparent that the
 lattice generalization immediately suggests itself.
 To be more precise, having taken into account the correspondence
 between (3) and (9), we propose the following general formula for
 spin matrix elements on  finite periodic lattice:
  $${ _{NS}\langle p_1,\ldots,p_K|\sigma|
 q_1,\ldots,q_L\rangle_{R}}=
 \sqrt{\xi \,\xi_T} \prod\limits_{i=1}^{K}\frac{e^{\nu(p_i)/2}}{\sqrt{N \sinh \gamma(p_i)}}
 \prod\limits_{j=1}^{L}\frac{e^{-\nu(q_j)/2}}{\sqrt{N \sinh
 \gamma(q_j)}}\times$$
 $$\times
 \prod\limits_{1\leq i<j\leq K}\frac{\sin \frac{p_i-p_j}{2}}
 {\sinh \frac{\gamma(p_i)+\gamma(p_j)}{2}}
 \prod\limits_{1\leq i<j\leq L}\frac{\sin \frac{q_i-q_j}{2}}
 {\sinh \frac{\gamma(q_i)+\gamma(q_j)}{2}}
 \prod\limits_{\substack{1\leq i\leq K \\ 1\leq j\leq L}}
 \frac{\sinh \frac{\gamma(p_i)+\gamma(q_j)}{2}}{\sin
 \frac{p_i-q_j}{2}}, \eqno(12) $$
 where $\xi_T$, $\nu(q)$ are defined by (4) and (5).
 All previous results can be easily obtained from this expression
 in appropriate limits. However, we have not yet found a
 rigorous proof of (12). Instead, since this formula should hold even
 on the finite lattice, we have verified it explicitly for
 small $N$.

 As an illustration, let us consider 3-row Ising chain in
 the ferromagnetic region of temperature parameter ($s>1$). In this
 case NS (R) momenta take on the values $\pi$, $\pi/3$, $-\pi/3$
 ($\,0$, $2\pi/3$, $-2\pi/3$). Each state contains
 either two particles or no particles at all. Since the integrals (4) and (5)
 can be alternatively written as
 \normalsize
 $$ \xi_T^{\;4}=\frac{{\prod\limits_q}^{(R)}{\prod\limits_p}^{(NS)}
 \sinh^2\frac{\gamma(q)+\gamma(p)}{2}}{{\prod\limits_q}^{(R)}{\prod\limits_p}^{(R)}
 \sinh\frac{\gamma(q)+\gamma(p)}{2}{\prod\limits_q}^{(NS)}{\prod\limits_p}^{(NS)}
 \sinh\frac{\gamma(q)+\gamma(p)}{2}}, \;\;\;
 \nu(q)=\ln \frac{{\prod\limits_p}^{(NS)}
 \sinh\frac{\gamma(q)+\gamma(p)}{2}}{{\prod\limits_p}^{(R)}
 \sinh\frac{\gamma(q)+\gamma(p)}{2}}\, ,$$ \normalsize
 then to verify (12) it suffices to prove ten relations:
 \small
 \begin{eqnarray} \nonumber { _{NS}\left\langle
       \emptyset|\sigma|\emptyset\right\rangle_{R}}^2 &=&
 \frac{\sinh \frac{\gamma_0+\gamma_{\pi/3}}{2}\sinh \frac{\gamma_{\pi}+\gamma_{2\pi/3}}{2}
       \sinh^2 \frac{\gamma_{\pi/3}+\gamma_{2\pi/3}}{2}}
       {\sinh\gamma_{2\pi/3}\sinh\gamma_{\pi/3}\sinh \frac{\gamma_0+\gamma_{2\pi/3}}{2}
       \sinh \frac{\gamma_{\pi}+\gamma_{\pi/3}}{2}}\; , \\
 \nonumber { _{NS}\left\langle -\pi/3,\pi/3|\sigma|
 2\pi/3,-2\pi/3\right\rangle_{R}}^2 &=&
 \frac{\sinh \frac{\gamma_0+\gamma_{2\pi/3}}{2}\sinh \frac{\gamma_{\pi}+\gamma_{\pi/3}}{2}
       \sinh^2 \frac{\gamma_{\pi/3}+\gamma_{2\pi/3}}{2}}
 {9\sinh\gamma_{2\pi/3}\sinh\gamma_{\pi/3}\sinh \frac{\gamma_0+\gamma_{\pi/3}}{2}
       \sinh \frac{\gamma_{\pi}+\gamma_{2\pi/3}}{2}}\; ,\end{eqnarray}
 \begin{eqnarray} \nonumber { _{NS}\left\langle
  \emptyset|\sigma|2\pi/3,-2\pi/3\right\rangle_{R}}^2
 &=&\frac{\sinh \frac{\gamma_0+\gamma_{\pi/3}}{2}\sinh \frac{\gamma_0+\gamma_{2\pi/3}}{2}}
 {12\sinh\gamma_{2\pi/3}\sinh\gamma_{\pi/3}
       \sinh \frac{\gamma_{\pi}+\gamma_{\pi/3}}{2}\sinh \frac{\gamma_{\pi}+\gamma_{2\pi/3}}{2}
       \sinh^2 \frac{\gamma_{\pi/3}+\gamma_{2\pi/3}}{2}} \; , \\
 \nonumber { _{NS}\left\langle -\pi/3,\pi/3|\sigma|\emptyset
   \right\rangle_{R}}^2 &=&
 \frac{\sinh \frac{\gamma_{\pi}+\gamma_{\pi/3}}{2}
       \sinh \frac{\gamma_{\pi}+\gamma_{2\pi/3}}{2}}
 {12\sinh\gamma_{2\pi/3}\sinh\gamma_{\pi/3}\sinh   \frac{\gamma_0+\gamma_{\pi/3}}{2}
 \sinh \frac{\gamma_0+\gamma_{2\pi/3}}{2}\sinh^2
 \frac{\gamma_{\pi/3}+\gamma_{2\pi/3}}{2} }\; ,\end{eqnarray}
 \begin{eqnarray}
 \nonumber{ _{NS}\left\langle \emptyset|\sigma|0,2\pi/3\right\rangle_{R}}^2=
    { _{NS}\left\langle \emptyset|\sigma|0,-2\pi/3\right\rangle_{R}}^2
    &=&
    \frac{1}  {12 \sinh \frac{\gamma_0 +\gamma_{\pi}}{2}
 \sinh \frac{\gamma_0+\gamma_{\pi/3}}{2}\sinh
 \frac{\gamma_{\pi}+\gamma_{\pi/3}}{2}\sinh\gamma_{\pi/3}}\; ,\\
 \nonumber  { _{NS}\left\langle -\pi/3,\pi|\sigma|\emptyset\right\rangle_{R}}^2=
 { _{NS}\left\langle \pi/3,\pi|\sigma|\emptyset\right\rangle_{R}}^2 &=&
 \frac{1} {12\sinh \frac{\gamma_0+\gamma_{\pi}}{2}\sinh
   \frac{\gamma_{0}+\gamma_{2\pi/3}}{2}
\sinh\gamma_{2\pi/3}\sinh \frac{\gamma_{\pi}+\gamma_{2\pi/3}}{2}}\; ,\end{eqnarray}
 \begin{eqnarray} \nonumber{ _{NS}\left\langle -\pi/3,\pi/3|\sigma|0,2\pi/3\right\rangle_{R}}^2=
 { _{NS}\left\langle -\pi/3,\pi/3|\sigma|0,-2\pi/3\right\rangle_{R}}^2
 &=&
 \frac{4\sinh \frac{\gamma_0+\gamma_{\pi/3}}{2}\sinh \frac{\gamma_{\pi}+\gamma_{\pi/3}}{2}}
 {9\sinh\gamma_{\pi/3}\sinh \frac{\gamma_0+\gamma_{\pi}}{2}}\; , \\
 \nonumber { _{NS}\left\langle -\pi/3,\pi|\sigma|
 2\pi/3,-2\pi/3\right\rangle_{R}}^2=
 { {_{NS}\left\langle \pi/3,\pi|\sigma|
 2\pi/3,-2\pi/3\right\rangle_{R}}}^2 &=&
 \frac{4\sinh \frac{\gamma_0+\gamma_{2\pi/3}}{2}\sinh \frac{\gamma_{\pi}+\gamma_{2\pi/3}}{2}}
 {9\sinh\gamma_{2\pi/3}\sinh \frac{\gamma_{0}+\gamma_{\pi}}{2}}\; ,\end{eqnarray}
 \begin{eqnarray}
 \nonumber { _{NS}\left\langle -\pi/3,\pi|\sigma|0,2\pi/3\right\rangle_{R}}^2=
 { _{NS}\left\langle \pi/3,\pi|\sigma|0,-2\pi/3\right\rangle_{R}}^2 &=& \frac19\; ,\\
 \nonumber { _{NS}\left\langle -\pi/3,\pi|\sigma|0,-2\pi/3\right\rangle_{R}}^2=
 { _{NS}\left\langle \pi/3,\pi|\sigma|0,2\pi/3\right\rangle_{R}}^2 &=&
 \frac49\; .\end{eqnarray}
 \normalsize
 This indeed can be done --- with a little bit cumbersome but
 straightforward calculation\footnote{Useful ``building blocks'', that greatly
   simplify it, can be found in the Appendix of \cite{ufz}.}. We have
 performed a similar check for small $N$ up to $N=4$ and we have no
 doubt in the validity of (12) for arbitrary~$N$. The rigorous proof of this formula
 will complete, in a sense, the study of the 2D Ising model in zero field.

 \begin{center}-----------------------------------\end{center}

 \vspace{0,5cm}  We thank B.~Banos, V.~N.~Rubtsov and V. N. Shadura
 for help and numerous stimulating discussions. This work was supported
 by the INTAS program under grant INTAS-00-00055.
 

\begin{thebibliography}{100}
  \bibitem{karowski} B. Berg, M. Karowski, P. Weisz, \textit{Construction of
    Green's functions from an exact S matrix},
    Phys. Rev. \textbf{D19}, (1979), 2477--2479.

 \bibitem{bugrij1} A. I. Bugrij, \textit{Correlation function of the
 two-dimensional Ising model on the finite lattice.~I}, Theor.
 Math. Phys. \textbf{127}, (2001), 528--548; \texttt{hep-th/0011104}.

  \bibitem{bugrij} A. I. Bugrij, \textit{Form factor representation of the
 correlation function of the two dimensional Ising model on a
 cylinder}, in \textit{Integrable Structures of Exactly
 Solvable Two-Dimensional Models of Quantum Field Theory}, eds.
 S.~Pakuliak and G. von Gehlen, NATO Sci. Ser. II Math. Phys. Chem. \textbf{35},
 Kluwer Acad. Publ., Dordrecht, (2001), 65--93;
 \texttt{hep-th/0107117}.

 \bibitem{we} A. I. Bugrij, O. Lisovyy, \textit{Magnetic susceptibility of
   the two-dimensional Ising model on a finite lattice},  JETP~\textbf{94}, (2002), 1140--1148
   [Zh. Eksp. Teor. Fiz.   \textbf{121}, (2002),
   1328--1338];
   \texttt{hep-th/0106270}.

 \bibitem{ufz} A.I. Bugrij, O. Lisovyy, \textit{Correlation function in
   the 2D Ising model on a cylinder}, Ukr. J. Phys. \textbf{47}, (2002), 179--196.

  \bibitem{fonseca} P. Fonseca, A. Zamolodchikov, \textit{Ising field theory
 in a magnetic field: analytic properties of the free energy},
 J. Stat. Phys. \textbf{110}, (2003), 527--590; \texttt{hep-th/0112167}.


 \bibitem{huang} K. Huang, \textit{Statistical mechanics}, 2nd ed.,
   Wiley, (1987).

 \bibitem{mccoy} B. M. McCoy, \textit{The connection between statistical
   mechanics and quantum field theory},
 in \textit{Statistical Mechanics and Field Theory}, eds. V.V Bazhanov
 and C.J. Burden, World Scientific, (1995), 26--128; \texttt{hep-th/9403084}.

 \bibitem{montroll} E. W. Montroll, R. B. Potts, J. C. Ward,
   \textit{Correlations and spontaneous magnetization of the two-dimensional
   Ising model}, J. Math. Phys. \textbf{4}, (1963), 308--322.

 \bibitem{palmer} J. Palmer, C. A. Tracy, \textit{Two-dimensional
   Ising correlations: convergence of the scaling limit},
   Adv. Appl. Math.~\textbf{2}, (1981), 329--388.

 \bibitem{smj} M. Sato, T. Miwa, M. Jimbo, \textit{Holonomic quantum
   fields. IV}, Publ. RIMS, Kyoto Univ. \textbf{15}, (1979), 871--972.

 \bibitem{yamada} K. Yamada, \textit{On the spin-spin correlation function in
   the Ising square lattice and the zero field susceptibility},
   Prog. Theor. Phys. \textbf{71}, (1984), 1416--1418.

 \end{thebibliography}
\end{document}